\documentclass[10pt,twocolumn,letterpaper]{article}

\usepackage{cvpr}
\usepackage{times}
\usepackage{epsfig}
\usepackage{graphicx}
\usepackage{amsmath}
\usepackage{amssymb}
\usepackage{tikz}
\usepackage{subcaption}
\usepackage{stackengine}
\usepackage{float}
\usepackage[breaklinks=true,bookmarks=false]{hyperref}

\cvprfinalcopy

\begin{document}

%%%%%%%%% TITLE
\title{Controllable Confidence-Based Image Denoising}

\author{Haley Owsianko \quad Florian Cassayre \quad Qiyuan Liang\\
École Polytechnique Fédérale de Lausanne ‐ EPFL, Switzerland \\
{\tt\small \{haley.owsianko, florian.cassayre, qiyuan.liang\}@epfl.ch} 
}

\maketitle

\begin{abstract}
Image denoising is a classic restoration problem. Yet, current deep learning methods are subject to the problems of generalization and interpretability. To mitigate these problems, in this project, we present a framework that is capable of controllable, confidence-based noise removal. The framework is based on the fusion between two different denoised images, both derived from the same noisy input. One of the two is denoised using generic algorithms (e.g. Gaussian), which make few assumptions on the input images, therefore, generalize in all scenarios. The other is denoised using deep learning, performing well on seen datasets. We introduce a set of techniques to fuse the two components smoothly in the frequency domain. Beyond that, we estimate the confidence of a deep learning denoiser to allow users to interpret the output, and provide a fusion strategy that safeguards them against out-of-distribution inputs. Through experiments, we demonstrate the effectiveness of the proposed framework in different use cases.
\end{abstract}

\section{Introduction}
Image noise removal is one of the fundamental problems in image restoration. The degradation model can be expressed as 
\begin{equation}\label{eqn:image_formation}
    y=x+n,
\end{equation} 
where $y$ is the noisy image captured by the camera, $x$ is the clean scene, and $n$ is the additive noise. One common assumption is that the $n$ is the additive white gaussian noise (AWGN). The goal of the image denoising is to recover the clean image $x$ from a noisy input $y$. Yet, they are inherently ill-posed problems as the plausible solutions highly depend on prior assumptions and are rarely universal. In addition, the measurement of the quality of a restoration algorithm also depends on the chosen assessment method \cite{quality_assessment}. Until now, there is no single metric that concludes with certainty the quality of the denoising output.

\makeatletter{\renewcommand*{\@makefnmark}{}
\footnotetext{Source code available under: \url{https://gitlab.epfl.ch/owsianko/controllable-denoising}}\makeatother}

Traditional algorithms focus on developing image priors to add constraints to the noisy image \cite{bm3d,WNNM}. With the blooming of deep learning (DL) and the creation of large datasets, new solutions to these problems have been developed \cite{dncnn,mwcnn,noise2noise,realimagedenoising}, outperforming the traditional ones quantitatively and qualitatively. However, nothing is perfect and DL methods have their own drawbacks: poor generalization on unseen data, e.g. images from another domain, and images with different noise levels \cite{SFM}. Also, deep networks are often treated as a black box for which it is difficult to evaluate the predictability and interpret the results. At times when deep learning makes mistakes that users are not able to identify, this motivates the need of having some control over the system and the ability to reason about the output \cite{confidence_classification_1}.

\begin{figure*}[t]
\begin{center}
\includegraphics[width=0.9\linewidth]{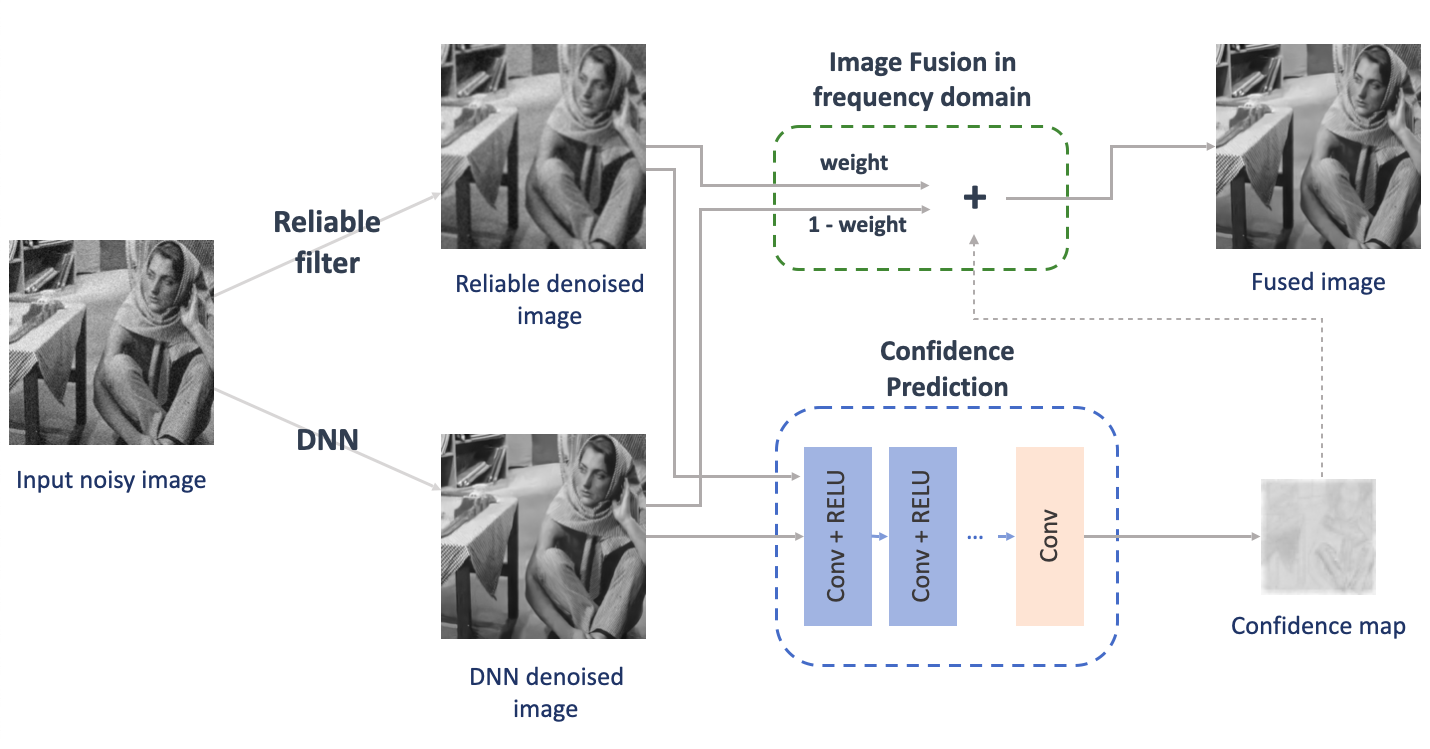}
\end{center}
\caption{Controllable confidence-based image denoising pipeline. The noisy image goes through both reliable filters and deep neural networks (DNN) to obtain two denoised images. The denoised image of DNN is further fed into the confidence prediction network to get a coarse-grained confidence map. Users can choose to either apply the confidence map to guide fusion or display it solely to interpret the output of the DNN method.}
\label{fig:pipeline}
\end{figure*}

There are various approaches proposed to address the above issues. As for generalization, common approaches that applied to any DL methods are taken, including data augmentation, training on a larger dataset, dropout layer in DL model, and so on \cite{brownlee2018better}. Specific to image denoising, in SFM, the authors alleviated the overfitting problems of denoising by masking the high-frequency component of the training images \cite{SFM}. Kim et al. \cite{denoising_overfitting} proposed to use an adaptive instance normalization during training together with a transfer learning schema to improve the robustness of the network. Regarding the interpretability, users essentially require more information than single output values to understand the mechanism, in particular, how confident the network is about its prediction. Even when dealing with unseen types of data, DL methods work as usual and give a normal output value, without any alert for the users. This limits its usage in many safety-critical domains. Some literature has proposed ways to model the confidence of an output \cite{confidence_classification_0,confidence_classification_1,confidence_classification_2}, either to regularize the prediction values or to detect the less confident samples directly. In \cite{uncertainty_deraining_0,uncertainty_deraining_1}, the authors incorporated the idea of a confidence map in deraining tasks to guide the residual map generation. The BIGPrior decoupled the hallucination and data fidelity component of the denoised output, and the final clear image is based on the confidence of the prior assumption \cite{bigprior}.

In this project we propose a new framework, controllable confidence-based image denoising (CCID), to address the issues with generalization and interpretability of the image denoising task, based on two observations. First, building a model that generalizes to all noise settings is now technically infeasible. However, differentiation between in-distribution data and out-of-distribution (OOD) data could be done. Second, convolution output on the original noisy images is reliable. Unlike DL methods that might make mistakes sometimes, low-pass filters (e.g., Gaussian filter) are always trustworthy. Although the filtered image does lose some details of the scene, there are never wrong hallucinations. As shown in Figure \ref{fig:pipeline}, CCID first processes the input noisy image to two intermediate denoised output. Next, the confidence prediction network evaluates the output of DNN to produce a coarse-grained confidence map. In the presence of three auxiliary images, namely the reliable denoised image, the DNN denoised image, and the confidence map, users control the fusion to produce the final denoised image. Our contributions can be summarized as follows. We propose and evaluate fusion methods in the frequency domain between the reliable denoised image and the DNN denoised image. We add the interpretability of DL methods through the confidence map. Confidence prediction identifies regions where the DL denoising method might make mistake. We give users the flexibility to smoothly vary from the reliable denoised image to the DNN denoised image. In the presence of mispredictions of DNN, users have the possibility to switch to the reliable denoised image. It gives CCID robustness against various kinds of data inputs. In addition, users can choose to use the confidence map as guidance to the fusion process and achieve better fusion quality. Finally we demonstrate the possibility of extending this approach to other restoration tasks such as super-resolution. The following sections are managed as follows. In Section \ref{literature_review}, we present a brief survey of related topics. Section \ref{implementation} introduces our detailed implementations. Section \ref{results} evaluate our proposed methods from different perspectives. Finally, Section \ref{conclusion} summarize the report and present future potentials.

\section{Literature review}\label{literature_review}

\subsection{Image denoising}
Before the era of deep learning, image denoising was based on prior knowledge of noisy images. BM3D \cite{bm3d} made the assumption that there are similar patches inside the same image. Collaborative filters were then used on every group of similar patches to denoise the image. In WNNM \cite{WNNM}, the researchers assumed that images could be represented as low-rank matrices. Taking the physical meanings of the singular value of the images, it achieved relatively good denoising results. Nevertheless, although these assumptions hold for common scenarios, the performance degrades when they do not. These methods require a known noise level, limiting their usage. Moreover, they generally attempt to solve complex optimization problems, which makes the denoising procedure computationally expensive. 

With the increase of computational power, DL methods became dominant. With a carefully designed architecture and a relatively large dataset, DL achieves superior performance compared to traditional algorithms. In DnCNN \cite{dncnn}, Zhang et al. observed that direct prediction of the denoised image from scratch is relatively hard for DL models. Instead, they proposed to predict the residual image, the noise $n$, from the input. Together with batch normalization, it turned out to vastly boost the performance in both accuracy and speed. The network also applies for super-resolution (SR) and JPEG blocking with a properly configured training set. MWCNN \cite{mwcnn} introduced a new type of downsampling and upsampling layer in building DL models through the discrete wavelet domain. Feeding the network with distinguished low and high-frequency information helps to guide the learning process, therefore, improving the performance. To boost the denoising performance further, the DDFN \cite{deep_boosting} integrated a dilated convolutional network with the deep boosting framework. Also, the generative adversarial network (GAN) \cite{GAN_denoising} is utilized to create virtual samples, augmenting the training set. FFDNet \cite{FFDNet} used a tunable noise level map as the input to tackle blind noise settings. BUIFD \cite{blind_bayesian} presented a blind and universal denoiser for AWGN removal to improve the model's generalization strength.

\subsection{Image fusion}
Image fusion is a classic problem in image processing. With the growth of current image processing applications, it is necessary to have a good fusion algorithm to integrate the complementary information from two or more images into one with improved quality and important features. In the literature, image fusion usually refers to integrate complementary information from multiple sensors, multiple time domains, and multiple views \cite{fusion_survey}. In this project, we focus specifically on fusion of the same scene but restored images using different methods. Classic image fusion techniques can be categorized into two groups, spatial and frequency domain. In general, direct fusion in spatial domain introduce some undesirable distortions in details of the image \cite{image_fusion_review}. Therefore, producing inferior results compared to fusion in the frequency domain. There are two widely used fusion techniques in the frequency domain, namely Discrete Cosine Transform (DCT) and Discrete Wavelet Transform (DWT) \cite{img_fusion_ref_0,img_fusion_ref_1,img_fusion_ref_2}. A 2D DCT transform $F(u,v)$ of an image $f(i,j)$ of size $N \times M$ can be modeled as
\begin{small}
\begin{align}
F(u,v) =& 
a(u)a(v)
\sum_{i=0}^{N-1}
\sum_{j=0}^{M-1}\nonumber\\
&f(i,j)
cos\left[ \frac{\pi u}{2N}(2i+1) \right]
cos\left[ \frac{\pi v}{2M}(2j+1) \right]. \label{eqn:dct}
\end{align}
\end{small}
Whereas for DWT, it is computed by applying a low pass filter $g$ and a high pass filter $h$ on the image pixels. The outputs contain the approximation coefficients (from the low-pass) and the detail coefficients (from the high-pass filter). The 2D DWT can be viewed as a combination of two 1D DWT. In other words, a horizontal 1D DWT on each row and a vertical 1D DWT on each column. The approximation coefficient $y_{low}$ can be computed as \begin{equation}
    y_{low}[n] = (\sum_{k=-\infty}^{\infty}x[k]g[2n-k])\downarrow_2,
\end{equation} where $\downarrow_2$ refers to downsampling of the output coefficients by 2. The detail coefficients are computed by replacing the low pass filter $g$ with the high pass filter $h$. The process can be repeated further to decompose the approximation coefficients with an increased frequency resolution.

\subsection{Confidence on DL}
The overconfident prediction issues of DL have limited its practical values. The authors of \cite{confidence_classification_0} pointed out that, although modern neural networks achieve high accuracy, they also generate miscalibrated outputs. The problem can be viewed from two angles. On the one hand, DL is unlikely to make confident mispredictions when evaluating inputs that are similar to the training set. On the other hand, when encountering unseen data, previously learned parameters hardly make sense. Even though DNN outputs valid answers, they are just unreliable at all. This kind of problem refers to as OOD. To solve the problem, researchers introduced the idea of confidence. In other words, the models should know what it does not know. In \cite{confidence_classification_2}, researchers introduced a baseline to evaluate the confidence of DNN, which was based on the insight that correctly classified examples tend to have greater maximum softmax probabilities than erroneously classified OOD examples. \cite{confidence_classification_3} demonstrated that mispredicted in-distribution data can be used as a proxy for OOD samples to mitigate the cost of collecting new OOD datasets. \cite{confidence_classification_1} proposed the correctness ranking loss for train neural networks that are aware of the confidence. Confidence has been widely discussed in the field of classification. However, to the best of our knowledge, it has seen little use in the field of image restoration. Two recent works \cite{uncertainty_deraining_0,uncertainty_deraining_1} introduced architecture to incorporate confidence inside the neural network to guide the deraining process, and the confidence map was calculated at multiple levels to achieve higher accuracy.

\subsection{Super resolution}
Single image super-resolution (SISR) is a restoration task that consists of reconstructing the high-resolution (HR) image from the low-resolution (LR) observations by adding missing pixel information. Given an LR image $y$, the image formation process can be modeled as
\begin{equation}
    y = (x \circledast k)\downarrow_s +\ n,
\end{equation}
where $x \circledast k$ denotes convolution between the HR image $x$ and unknown blurry kernel $k$, $\downarrow_s$ refers to the downsampling with a scale $s$, and $n$ is the independent noise. In reality, the $k$ is unknown, and one common yet unrealistic assumption is that the degradation kernel is bicubic. Depending on the source, the noise $n$ can be assumed to equal zero. SISR problem is similar to image denoising if we view $y$ in Equation \ref{eqn:image_formation} as the bicubic upsampled low-resolution image and $x$ as the high-resolution image. The reconstruction involves the removal of noise in $y$ to recover $x$. The usage of DL for this task is even more prominent than for denoising, and most approaches rely on deep residual networks. In RCAN \cite{rcan}, the authors proposed a residual in residual structure to train a very deep network efficiently and a channel attention mechanism to improve the results. In SAN \cite{san}, Dai et al. proposed an architecture capable of extracting interdependent information as second-order features. Liu et al. proposed HBPN \cite{hbpn}, a hierarchy of residual hourglass modules that aim to provide a better error estimation and achieve state-of-the-art performance. However, the generalization ability of these models is limited by the finite kernel set $k$ and the scaling factors $s$ \cite{blind_kernel_problem_sr}. Recent works incorporated the degradation kernel $k$ explicitly inside the networks to mitigate the problem \cite{blind_kernel_sol_sr_0,blind_kernel_sol_sr_1}.

\section{Implementation}\label{implementation}
As shown in Figure \ref{fig:pipeline}, our method consists of three major components: image denoising, confidence prediction, and image fusion. The noisy input image first goes through both a reliable filter and a deep denoising network, which results in two different denoised images. Next, the confidence prediction network takes the residual map by DNN method, the original noisy image, and reliable filtered denoised image as inputs, and outputs a coarse-grained confidence map which is eight times smaller in both width and height compared to the original input. With the presence of the confidence map, users have a better idea about in which areas that deep denoising network might go wrong, and control the weighted fusion between two different denoised images. If users observe that the DNN method is highly unconfident, they could put more weight on the reliable filters, resulting in a more pleasant denoised scene. We proposed two modes of fusion. One is unguided fusion, which does not utilize the information from the confidence map. The other is the confidence-aware fusion, i.e. using the confidence map as auxiliary information to refine fusion results. In the following subsections, we discuss each component in detail regarding the implementation. We also offer a graphical user interface (GUI) to simplify user interaction.  It comes as the last part of this section.

\subsection{Image denoising}

\subsubsection{Reliable denoising}
We define reliable denoising as a method that makes few prior assumptions to provide a highly predictable output improving the noise removal. The few assumptions on these methods give them high robustness against different noise settings. The most straightforward example of such a method is the gaussian filter, acting as a low-pass filter and therefore alleviating the noise through averaging. As a side-effect, the image loses some of its sharpness due to blurring. We also provide support for other filters such as bilateral filtering and non-local means (NLM). For the experiments described at the end, we resorted to using the gaussian filter, as it outputs consistent results and does not require tweaking parameters other than the kernel size.

\subsubsection{DNN denoising}
In general, DL methods for denoising show superior performances compared to traditional methods and, of course, reliable filters. However, it is in the settings that the DNN has seen similar data during the training phase. The selected training set serves as the prior knowledge that the DL methods have. Their major strength and weakness at the same time is their ability to produce highly plausible results, but for which the information in the original noisy image could be distorted. This effect is particularly noticeable on extreme super-resolution tasks, which is discussed later. We call it "hallucinatory" as it is very distinctive from averaging or interpolation, and often correlates to the learned prior information. In this project, our focus is not to develop a denoising architecture that out-performs the state-of-art. Instead, we would like to explore the generalization problem when the inputs are OOD. All current DL denoising methods suffer more or less from these issues. In the seek of simplicity, we adopted the well-known denoising model DnCNN \cite{dncnn} as the default model. To facilitate the exploration of the OOD phenomena, we had the DnCNN trained on $\sigma=25$ using 400 images from the Berkeley Segmentation Dataset (BSDS500) \cite{bsd}, where $\sigma$ is the noise level. The DnCNN outputs the residual map ($n$) from the original noisy image ($y$). The final clean image ($x$) is computed by $x = y - n$.

\subsection{Confidence prediction}

\subsubsection{Architecture}
\begin{figure}[t]
\begin{center}
\includegraphics[width=0.8\linewidth]{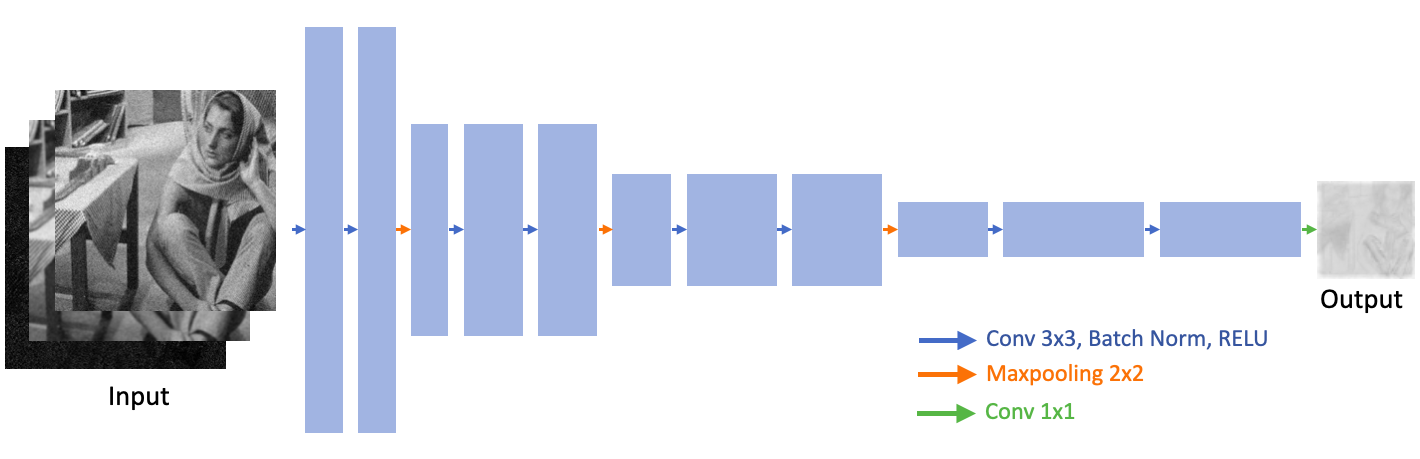}
\end{center}
\caption{The architecture of the confidence prediction network. The input consists of three images, the original noisy image, the reliable filtered denoised image, and the residual map from DNN. The output is a coarse-grained confidence map, eight times smaller in both width and height compared to the input image. The architecture consists of several layers of convolution and pooling.}
\label{fig:confidence_archi}
\end{figure}

The architecture of the confidence prediction network is shown in Figure \ref{fig:confidence_archi}. It takes the original noisy image, the reliable filtered denoised image, and the residual map from DNN as input. The intention is to evaluate whether the residual map is wrong or not. The original noisy image and the reliable filtered denoised image offer extra information and structural features to improve the convergence speed and accuracy. Our first attempt was predicting confidence value at the pixel level. However, there was too much fluctuation in the residual map. As a result, the prediction was far from accurate and unusable. The next attempt, which is the current model, is to predict a confidence value for every $8 \times 8$ region. In other words, the confidence map is $8$ times smaller in both width and height than the original image. The $8 \times 8$ region is a compromise between localizability and precision. The region-based approach produces fairly precise predictions and offers useful confidence information for users to identify regions that may contain incorrect hallucinations. The small-sized confidence map is achieved by multiple pooling layers. The accuracy relies on the convolution layers to extract useful features.

\subsubsection{Data generation}
As mentioned previously, the input would be the stacked array of the original noisy image, the reliable filtered denoised image, and the residual map, which is the difference between the original, noisy input, and the output of the DnCNN.

Since we only focused on gray scale image denoising, it results in a 3-channel input tensor. The intention is to evaluate the confidence of the DNN, namely how much error it makes. The ground truth $y_{confidence}$ is computed as 
\begin{equation}
    y_{confidence} = 1- 
    \frac{\left\lVert y_{clean\_image} - y_{dnn\_denoised} \right\lVert \downarrow_8}
    {\sigma_{max}}.
\end{equation}
$\left\lVert y_{clean\_image} - y_{dnn\_denoised} \right\lVert$ computes the $L_1$ norm of the difference between the clean image and the DNN denoised image. It is average downsampled by $8$ to get the error that DNN has made in a $8 \times 8$ region. The error is limited to $\sigma_{max}$. In our experiments, we only consider the noise range between $0$ and $100$ because images with noise levels over $100$ are almost unrecognizable. Therefore, $\sigma_{max}$ is set to be $100$.

For each image, a set of sub-images is generated. Those sub-images are referred to as patches and are then used to create the actual data points used for training the network, and each patch goes through standard data augmentation procedures to be more robust against overfitting. For each patch, a value $\sigma_p$ is sampled from a uniform distribution over $[0, 100]$, and is then used to generate the noise. This ensures that each noise level appears equally often in the dataset.

The data generated in this way is managed by a custom subclass of torch's Dataset. This class, ImageDataset, takes care of abstracting away the data generation and writing its results to disk to save computation time earlier. Additionally, data generation was written in a way that allows for GPU acceleration.

\subsubsection{Loss function}
\label{loss_function}
The confidence predictor intends to show users regions where DNN makes mistakes. Therefore, it is acceptable to have a lower predicted value compared to a higher one. The extreme case would be all regions have a high confidence value $1$. To suppress an overconfident confidence predictor, we adapted the loss function, the sum of the squared estimate of errors (SSE), to favor under-confident predictions by
\begin{align}
SSE_{confidence} =
& SSE[output < target]*p_{under} \nonumber\\
& + SSE[output \ge target]*p_{over},
\end{align}
where $p$ is the penalty coefficient. Through giving a higher $p_{over}$ than the $p_{under}$, we prevent it from being overconfident on evaluation to some extent.

\subsubsection{Training}
\begin{figure}
    \centering
    \scalebox{0.4}{\input{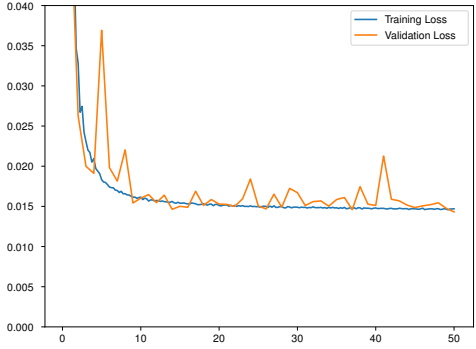}}
    \caption{The training and validation loss of the model. The x-axis stands for the number of training epoch.}
    \label{fig:losses}
\end{figure}

The network was trained using the standard procedure. Many values of hyper-parameters were tried. At first, the network was slightly overfitting. After we introduced a weight decay of $10^{-5}$, this problem was partially remedied. A value of $2 \cdot 10^{-4}$ produced better results. It did however also lead to a less stable validation loss curve. The final model was trained with a value of $10^{-4}$, which still produced a rather jittery validation loss. The plot of the training and validation loss can be seen in Figure.
\ref{fig:losses}.

\subsection{Image fusion}
In this part, we describe procedures to fuse a blurry but reliable signal with a sharp but hallucinatory one. We discuss the effectiveness of these fusion techniques. Two categories of tasks were considered. The first one is relatively general and consists of fusing the two images according to a single scalar parameter. The second relies on an additional confidence input.

\subsubsection{Unguided fusion}
\begin{figure}[t]
\begin{center}
\scalebox{0.5}{\input{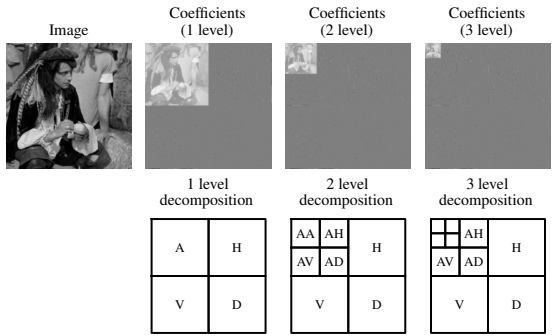}}
\end{center}
\caption{An illustration of 2D multilevel DWT on an image. A, V, H, D stand for the approximation, vertical, horizontal and diagonal wavelet coefficients respectively.}
\label{fig:dwt_fusion}
\end{figure}

For this approach, intuitively we wanted the two extreme values of this parameter to yield the two input images respectively, while the values in between would perform a combination of the two in a way that had to be defined. A scalar interpolation on the spatial domain provides an accurate fusion in terms of pixel-wise distance, but the resulting images are hardly plausible due to ghosting \cite{img_fusion_ref_2}. We adapted fusion techniques in the frequency domain to combine the structure of both images smoothly. Here we consider two strategies, one in the DWT domain and the other in the DCT.

As for DWT, to generate a smooth combination between the two input images, we considered a multilevel DWT. Figure \ref{fig:dwt_fusion} shows the DWT on an image from level 1 to level 3. A higher level indicates a higher frequency resolution. Two images of the same size go through the same level of the DWT, and coefficients at each level are retrieved accordingly. The fusion is then computed as the weighted average between the two. Besides, we proposed an advanced version to incorporate the cross-correlation of two coefficients to perform the fusion. The idea is that, if the reliable denoised image and the DNN one have a similar coefficient, there is no reason that we give too much weight to the possible wrong estimations.

For DCT, over a 2D domain, the spectral power is usually modeled as a probability density function of a Laplacian distribution; however it has been shown that it follows more closely a half-Gaussian probability density function \cite{dct_distribution}. Given this observation and some assessments we derived a mask-based technique, in the DCT domain:

\begin{align}
& M_w(\omega_x,\omega_y) = \exp(-(\omega_x^2+\omega_y^2)/(2s)) \nonumber\\
& \text{with } s = a\left(\frac{1}{1-w+\epsilon}-1\right).
\end{align}

$w \in [0, 1]$ is the provided fusion weight. The value $a$ controls the scale of the weight and was determined empirically. The value $\epsilon$ should be a small positive number.

The mask is then used to perform a simple interpolation on the values of each frequency, namely when $M=1$ the output equals the hallucinatory input and when $M=0$ it equals the reliable one. The low frequencies of the hallucinatory image are the first to get fused, followed by higher frequencies as the weight increases. For most image restoration tasks, including denoising and super-resolution, the low frequencies are the easiest to predict with high confidence - the reason why we integrate them first. An issue that we encountered in our previous attempts was that even though the fusion was smooth, it would introduce small wave artifacts near edges which were undesired. We suppose that it was due to an incorrect estimation of the spectral power over the frequency space, as it is absent from the results of this method.

\subsubsection{Confidence-aware fusion}
In the previous part, we discussed our initial strategies to fuse a DNN denoised output with a reliable component. In this part, we describe strategies that build on top of the previously defined functions and rely on the auxiliary confidence map component. The confidence map contains spatial reliability information that allows us to perform a fine-grained fusion. The goal is then to preserve the details when the risk of error is low, and otherwise average them to mitigate the consequences of an error. Remark that the confidence map is only an estimation and can be wrong too. However, as shown during the evaluation part this happens sufficiently rarely and the reliability can be provided to the user.

For DWT, we introduced patch-wise fusion. Patch-wise fusion refers to performing the DWT on each small $8 \times 8$ region, and stitch all fusion results together to get the whole image. Each $8 \times 8$ region contains a confidence value, therefore, we could use the confidence value on each region to guide the fusion individually. More precisely, we updated fusion weight on each region through
\begin{equation}\label{eqn:updated_conf}
    w_{region} = w \cdot (1 + c - t),
\end{equation}
where $w$ is the global fusion weight, $c$ is the confidence of that region, and $t$ is the threshold. The intuition is that if the confidence value is higher than the threshold, we give more weight to the DNN denoised image to obtain a better denoising result safely. In practice, the threshold is set to $0.8$ according to the data distribution of in-distribution data and OOD data, which is discussed later in Section \ref{results}.

Unlike DWT, which produces stable transformation even on an $8 \times 8$ region, DCT is incapable of adopting the same strategy as DWT. 
As the normal DCT fusion already works with a single global weight for which the scale is evenly spread, we chose to do a pixel-wise application of this function. One caveat is that each value on the confidence map represents an $8\times8$ pixels patch on the images. Therefore we rescale the confidence map using bicubic interpolation to match the input size and to minimize the artifacts. Also, following Equation \ref{eqn:dct}, in practice, it is not required to compute the full transform and inverse transform for every confidence value.

\subsection{User interface}

\begin{figure}[t]
\begin{center}
\includegraphics[width=\linewidth]{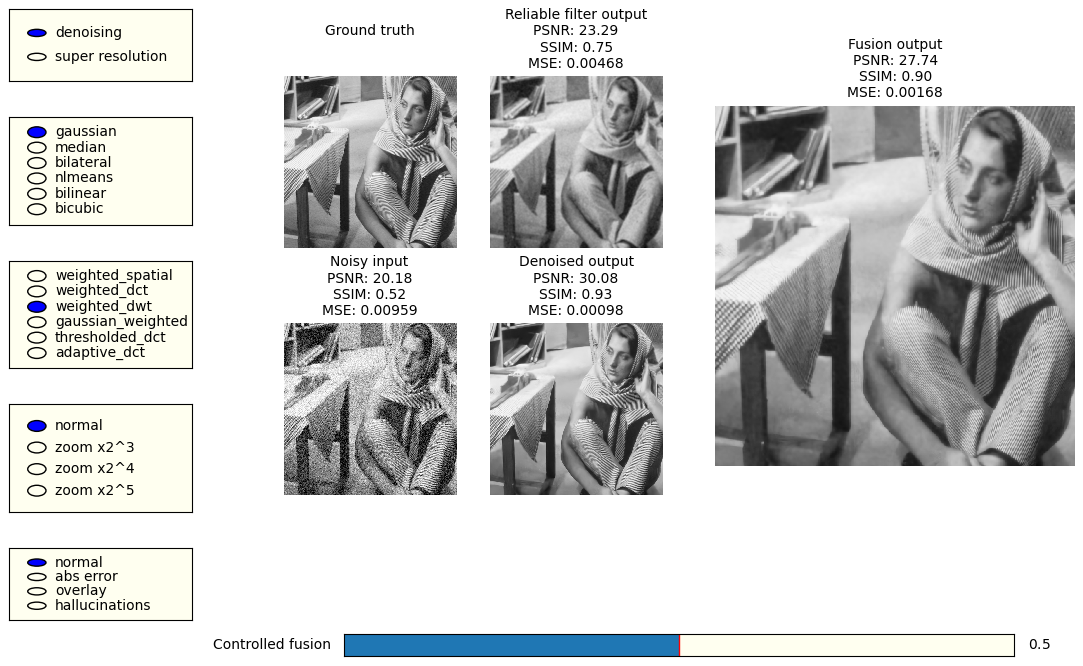}
\end{center}
\caption{Screenshot of the user interface. It is possible to switch between denoising and super resolution, select the reliable filter and fusion strategy, modify the zoom level and change the visualization mode (eg. highlighting the predicted hallucinations, or the error with respect to the ground truth). The ground truth is not available in evaluation mode.}
\label{fig:gui}
\end{figure}

Our project is fundamentally centered around users. One of the aspects is the input and corresponds to a specific need. For instance, we could assume that people working in medical imagery would be interested in having the possibility to configure the system to work at the designated confidence level, based on their requirements. On the other hand, the users must be able to interpret and reason about the output. Taking the same example, not only should it ease their job but it must absolutely reduce the risk of misinterpretation; as it is the very problem we are trying to solve.

To make our research process more effective, we designed early on a user interface (GUI) that provides interactivity (Figure \ref{fig:gui}). We kept in mind the two aspects while building this tool and investigated different strategies to present the information in a meaningful way.

\section{Results}\label{results}
In this section, we evaluate our proposed architecture from multiple perspectives both qualitatively and quantitatively. We show the flexibility of our controllable fusion strategy. With the guidance of the confidence map, users are able to interpret the results of DL methods and to make further decisions based on the results. Also, we evaluate the confidence-guided fusion algorithm. Last, we access the effects of OOD and the same infrastructure on SISR.

\subsection{Controllable fusion}
\begin{figure*}[h]
\captionsetup[subfigure]{labelformat=empty}
\begin{center}
\input{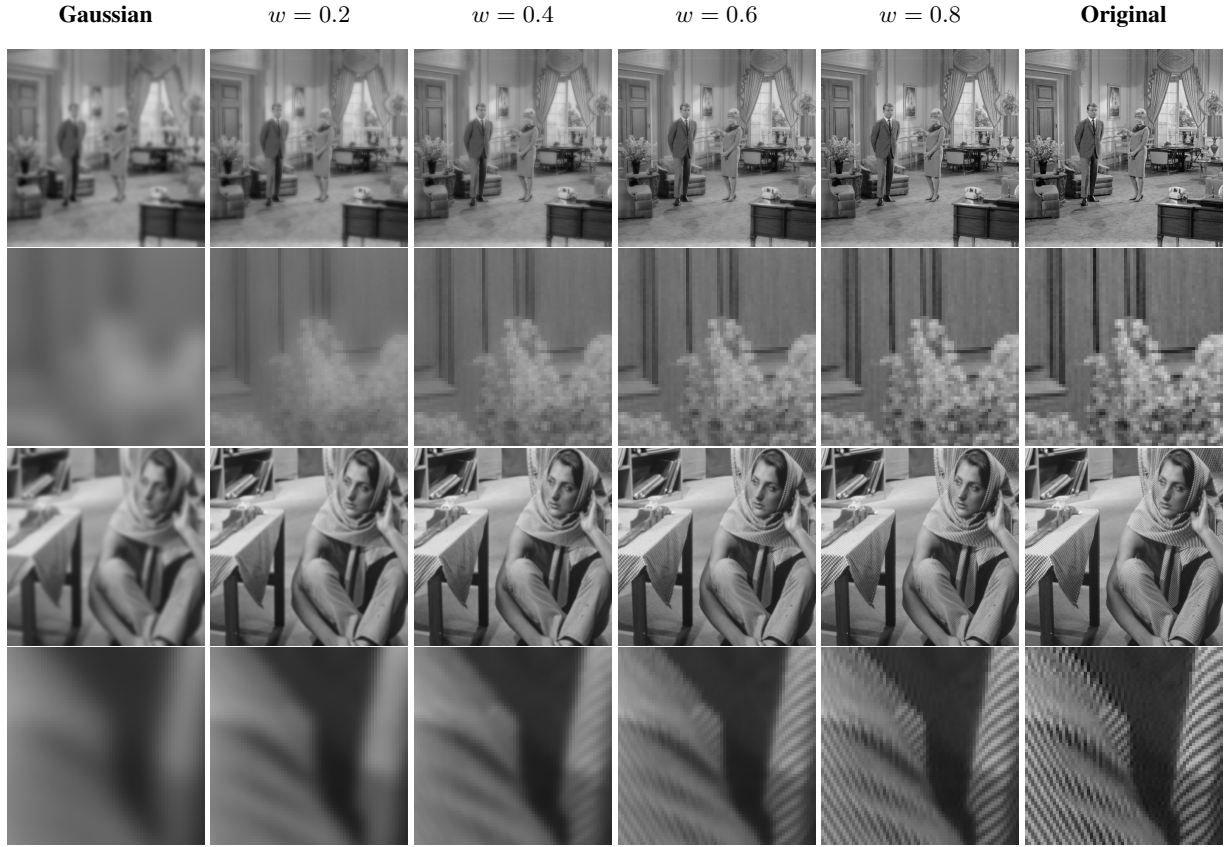}
\end{center}
\caption{Comparison of the fusion technique described above for different fusion weights $w$. The Gaussian filter used here has parameter $\sigma=4$ for exaggeration. The hallucinatory image is replaced by the ground truth for comparison purposes. First two rows represent the fusion results using DWT and last two rows using DCT. Observe that the edges gradually become less blurry.}
\label{fig:controllable_fusion}
\end{figure*}

Figure \ref{fig:controllable_fusion} shows the qualitative results of the proposed fusion algorithms, DWT and DCT. For the illustration purpose, we substitute the DNN denoised images with the clean images, and the Gaussian filter is applied to the clean images directly. As we could observe, the images vary smoothly from the Gaussian blurred one to the clean image, with gradual increasing details with the $w$ increasing. With the zoom-in details, it is easy to find out the edges and textures monotonically appear in the second and fourth row of the Figure \ref{fig:controllable_fusion}. With a controllable fusion weight $w$, users are free to choose a combination of the reliable filtered image and the DNN denoised images. For the in-distribution data, when DNN demonstrates predominant advantages, users can choose a higher $w$, up to $1$. When users are not satisfied with the DNN denoised images, they could switch to a lower $w$ to achieve the most pleasant results. Detailed use cases that DNN fails to produce good results are discussed in Section \ref{OOD}.

\begin{figure}
\begin{center}
\scalebox{0.2}{\input{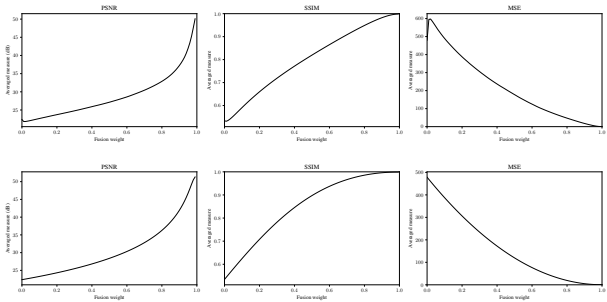}}
\end{center}
\caption{Quantitative evaluation of the DCT (first row) and DWT (second row) fusion performance with a gaussian filter $\sigma=4$, over the BSD Set68 images corpus.}
\label{fig:dct_quantitative}
\end{figure}

To support our assertions, we also evaluate the results using quantitative measures, including the peak-signal-to-noise ratio (PSNR), the structural similarity index measure (SSIM), and the mean squared error (MSE). The corresponding quantitative results of the same experiment settings as the Figure \ref{fig:controllable_fusion} are shown in Figure \ref{fig:dct_quantitative}. The gradual increased PSNR, SSIM, and decreased MSE are in accordance with our qualitative observations. DWT and DCT fusion methods achieve similar results with small variations for the chosen metrics.

\begin{figure*}[h]
\captionsetup[subfigure]{labelformat=empty}
\begin{center}
\input{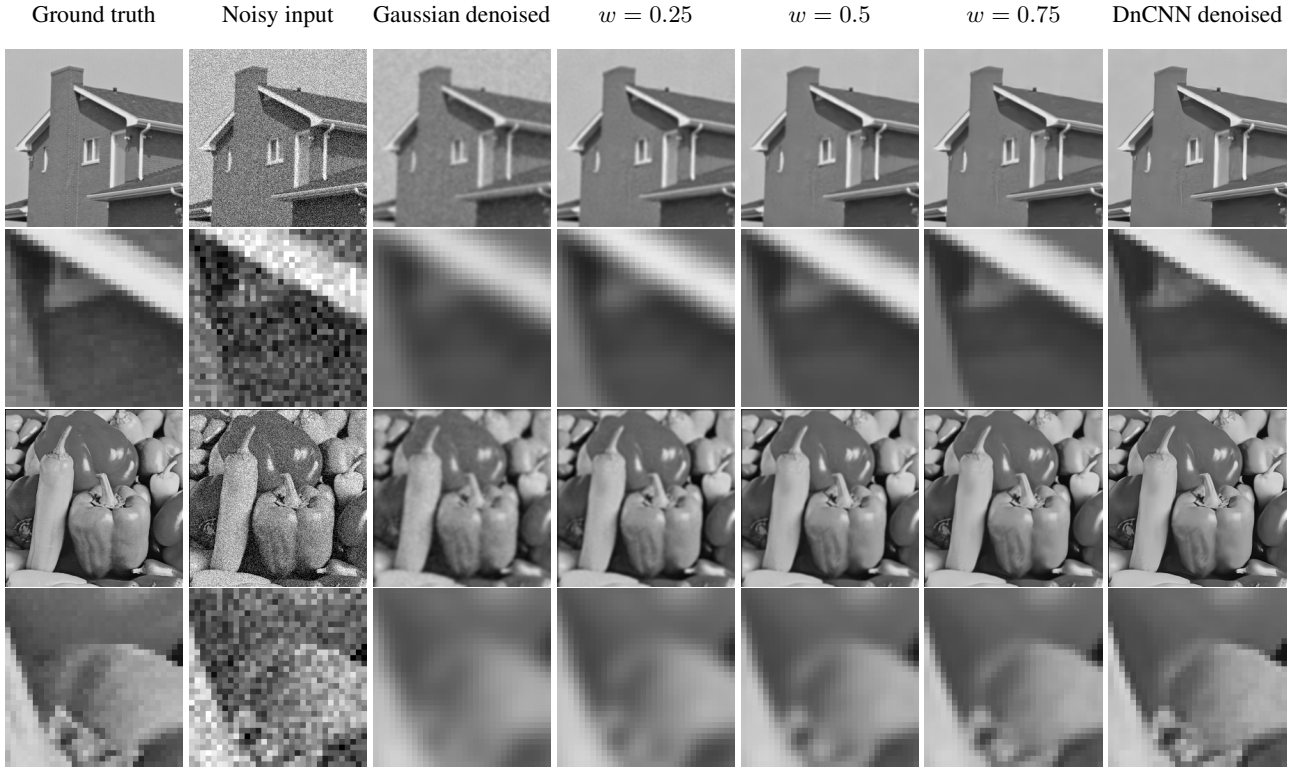}
\end{center}
\caption{Application of controllable fusion to denoising. The two images use the DCT algorithm on noisy inputs $\sigma=25$. The weight $w$ controls the fusion between the reliable image and the DNN output, as in Figure \ref{fig:controllable_fusion}.}
\label{fig:controllable_denoising}
\end{figure*}

\subsection{Confidence prediction}
After experimenting with the neural network, we ended with an SSE loss of around 0.0146 for the training set and 0.0142 for the validation set. The test set ended up at 0.0145. While it is odd that the training loss is higher than for the test and validation sets, the model performs rather well. All three sets contain an input with size $40 \times 40$, and an output with size $5 \times 5$. The SSE loss is indeed small and the prediction is quite close to the ground truth. Overfitting ended up minimal, as the two losses were very close together.

\begin{figure}[t]
    \centering
    \scalebox{0.20}{\input{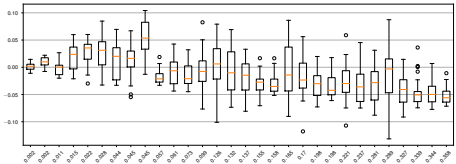}}
    \caption{The boxplot of the difference between individual pixels of target confidence and network output. A positive value indicates that the target confidence is larger than what the network predicted, meaning the the network is under-confident. A negative value means the network is over-confident.}
    \label{fig:confidence_stats_diff}
\end{figure}

\begin{figure}[t]
    \centering
    \scalebox{0.18}{\input{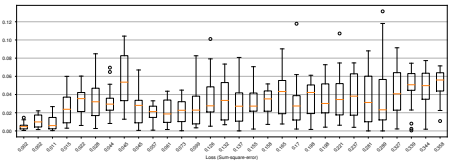}}
    \caption{The absolute value of the data presented in Figure \ref{fig:confidence_stats_diff}}
    \label{fig:confidence_stats_abs}
\end{figure}

However, an abstract number such as the SSE loss is rarely enough to assess how well a model is performing. It is worthwhile to look at the statistics corresponding to the loss. Figure \ref{fig:confidence_stats_diff} and Figure \ref{fig:confidence_stats_abs} show statistics for thirty different test images. The network still tends to be overconfident on this set. On both the training and the validation set, it has no problem being under-confident. This is perhaps due to the training and validation sets coming from a similar distribution. However, since the average loss is roughly the same, it is fair to assume the network trained well. Additionally, test samples with the largest losses tend to also have surprisingly little variance in absolute terms. In a way, this constitutes a success for our custom Loss Function, as defined in Section \ref{loss_function}. When the network overestimates the confidence, at least there is some guarantee of consistency. Figure \ref{fig:confidence_stats_abs} gives us insight in how effective the network is. For instance, we noticed that most of the items with low loss tend to not deviate much from the target, as expected. The vast majority of points have a loss of less than $0.05$, and thus Figure \ref{fig:confidence_stats_abs} and Figure \ref{fig:confidence_stats_diff} are fairly representative.

\begin{figure*}
\centering
\input{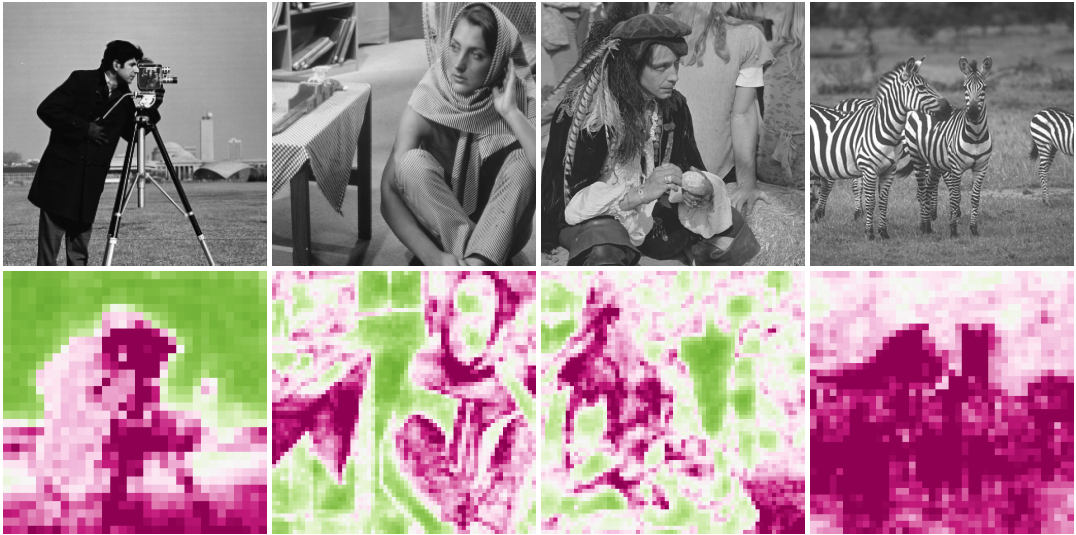}
\caption{Visualization of the confidence maps for $\sigma=25$ noisy inputs (only the ground truth is shown). The threshold is set at $0.95$, all the values above are colored green while the ones below are purple. The color intensity represents the distance to the boundary. The confidence map is $8$ times smaller than the input stack.}
\label{fig:confidence_interpolation}
\end{figure*}

Figure \ref{fig:confidence_interpolation} shows the predicted confidence map for each corresponding images. For textureless regions, e.g., sky, wall, and plain color T-shirts, the confidence is high. For regions with a lot of variations, e.g., grass ground, table cloth, and hairs, the confidence is relatively low. It follows the general assumption: the corruption of high-frequency components is significantly harder or impossible to recover. Therefore, high-frequency regions are likely to be places where DNN makes greater mistakes, which corresponds to a low confidence value.

\subsection{Fusion with confidence}
\begin{figure}
\begin{center}
\scalebox{0.32}{\input{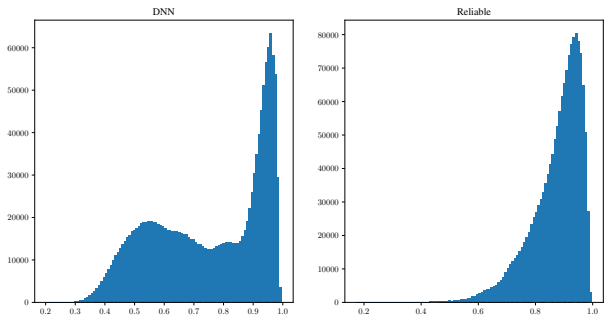}}
\end{center}
\caption{The distribution of the ground-truth confidence value using DNN and reliable filter respectively. The confidence value is negatively correlated to the mistakes that each method makes.}
\label{fig:confidence_distribution}
\end{figure}

In this part, we evaluated the confidence-guided fusion method. First, to understand confidence. Figure \ref{fig:confidence_distribution} plots the ground-truth confidence distribution using the DNN and the reliable filter respectively. Here, we use DnCNN trained on $\sigma=25$ and a Gaussian filter. As we could observe from the plot, the results of the Gaussian filter have an exponential distribution as expected when varying the noise level from $0$ to $100$, the distribution of the DNN method is kind of weird. The weirdness is also anticipated since the model is only trained on $\sigma=25$, and it does not generalize to other noise levels. We could observe a valley around confidence value equals to $0.8$. For the regions that have a confidence value below $0.8$, it is better to have more reliable components. This observation built up our updated confidence schema, as shown in Equation \ref{eqn:updated_conf}. The confidence threshold is then set up to $0.8$

\begin{figure*}
\captionsetup[subfigure]{labelformat=empty}
\begin{center}
\input{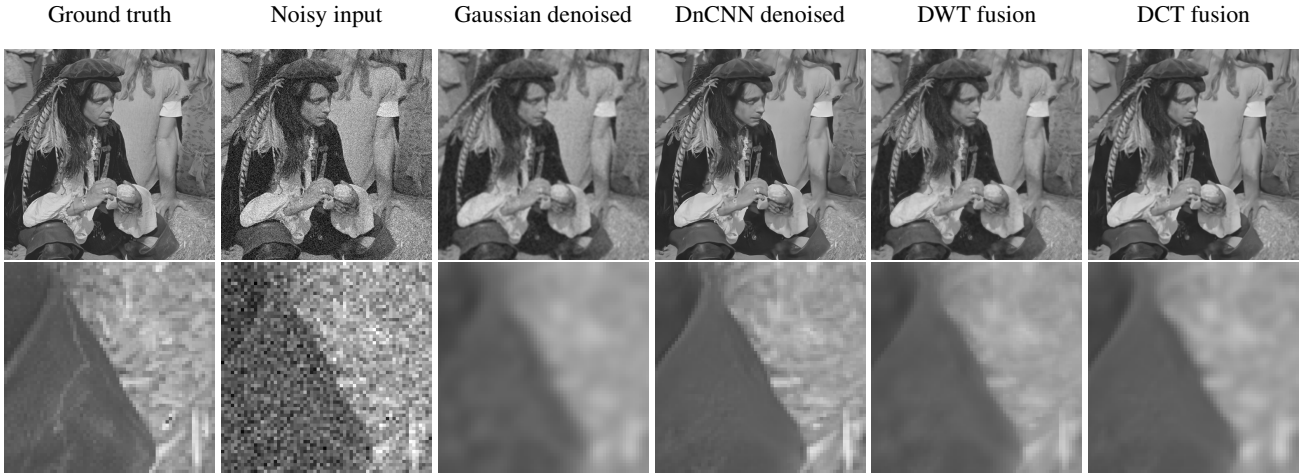}
\end{center}
\caption{Confidence aware DWT and DCT fusion for noise $\sigma=25$. On the magnified fragment of the image (bottom row), the left part was estimated to have higher confidence than the right one. For both techniques, the fused output is both sharp and reliable for the left part. The right part on the other hand was mostly contributed by the reliable input and is thus blurry. Note that the model was trained to be cautious rather than overconfident, which is the reason why there is bleeding on the transition.}
\label{fig:conf_guided_fusion}
\end{figure*}

\begin{figure}[t]
\begin{center}
\includegraphics[width=\linewidth]{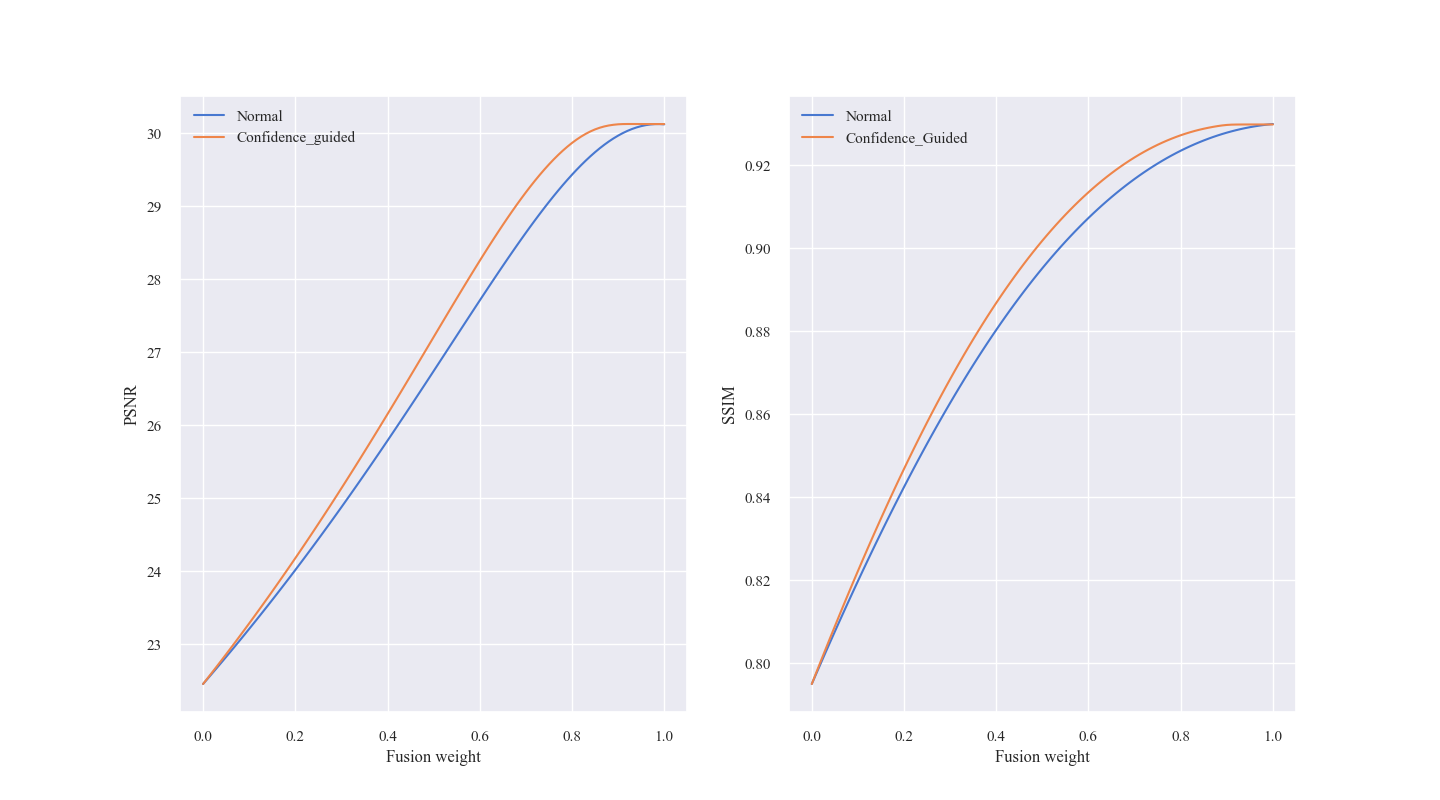}
\end{center}
\caption{PSNR and SSIM distribution comparison between confidence-guided fusion and normal fusion. Confidence-guided fusion has a slightly better statistical performance at every fusion weight.}
\label{fig:confidence_fusion_stat}
\end{figure}

Figure \ref{fig:conf_guided_fusion} shows the fusion results using confidence. The magnified regions (bottom row) show two areas with disparate confidence values, the plain cloth (left part) has a higher confidence value compared to the threshold, the right part contains a lot of textures, therefore has a lower confidence value. For the confidence fusion, we intentionally put a relatively higher weight on the confident region through scaling the global fusion weight using Equation \ref{eqn:updated_conf}. Compared to the fusion method without confidence under the same fusion weight, we could observe that the confidence-aware methods perform slightly better by preserving the trustworthy component from the DL method. The observation is also backed up by the statistical evaluation, as shown in Figure \ref{fig:confidence_fusion_stat}. Although the proposed method is not capable of improving the overall performance of the in-distribution denoised results. It is still worthwhile for offering better results at every fusion weight and giving users a higher possibility to select the best candidate. For in-distribution data, we almost always achieve the best performance at $w=1$ (DNN denoised image), and all mentioned fusion methods seem meaningless. However, it is not the case. As we would discuss in the following section, their effectiveness gets larger when dealing with OOD data, and in a real-life deployment, it is very likely that we are dealing with data that are not strictly similar to the training set.

\subsection{Out-of-distribution data}
\label{OOD}

\begin{figure*}
\captionsetup[subfigure]{labelformat=empty}
\begin{center}
\input{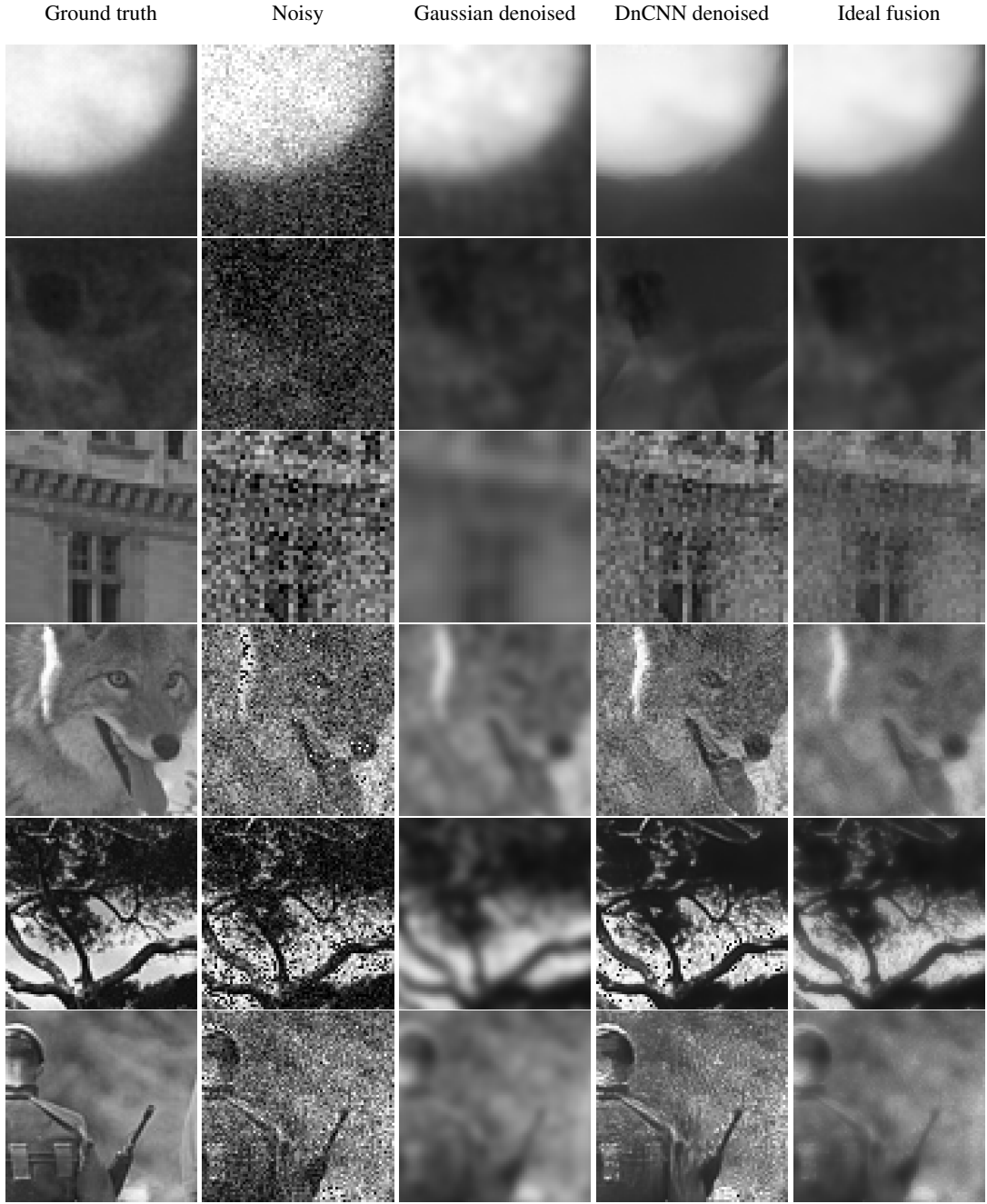}
\end{center}
\caption{Sample zoomed-in outputs of fusions that optimize the three measures (as given by the extremum point shown in Figure \ref{fig:ood}) on OOD data. Respective fusion weights were $0.55$, $0.25$, $0.5$, $0.2$, $0.4$, and $0.15$. Rows 1 and 2 are from the FMD dataset \cite{fmd_dataset} and have noise $\sigma=25$. Row 3 and 4 come from the BSD dataset with Gaussian noise $\sigma=35$. Row 5 and 6 come from the BSD dataset with Poisson noise having the same standard deviation as Gaussian noise $sigma=25$. Observe that in the fused output, the incorrect hallucinations introduced by DnCNN are mitigated in exchange for details.}
\label{fig:ood_fusion}
\end{figure*}

\begin{figure}[h]
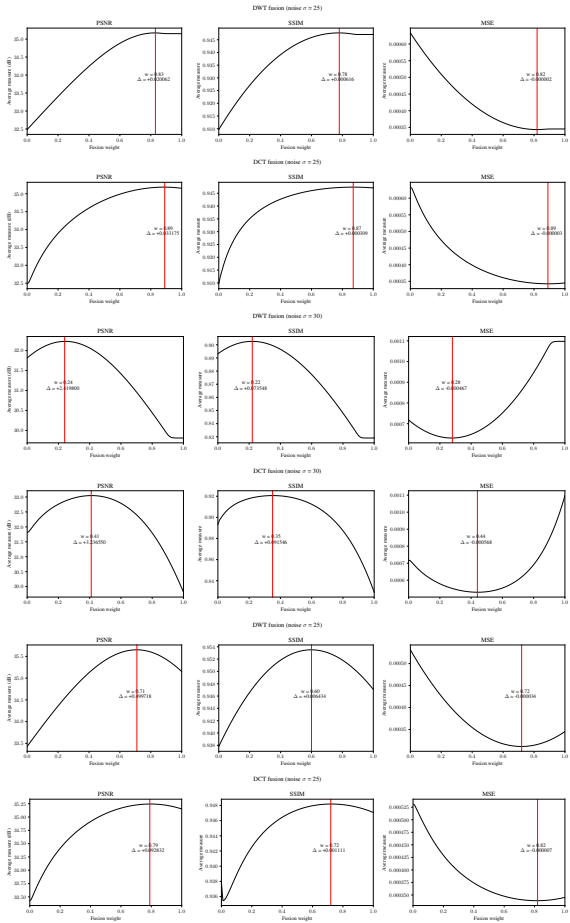

\begin{center}
\scalebox{0.2}{\input{images/OOD/ood_dwt_25.pgf}}
\scalebox{0.2}{\input{images/OOD/ood_dct_25.pgf}}
\scalebox{0.2}{\input{images/OOD/ood_dwt_30.pgf}}
\scalebox{0.2}{\input{images/OOD/ood_dct_30.pgf}}
\scalebox{0.2}{\input{images/OOD/ood_dwt_25_poisson.pgf}}
\scalebox{0.2}{\input{images/OOD/ood_dct_25_poisson.pgf}}
\end{center}
\caption{Row 1 and 2 are for noise $\sigma=25$ using DWT and DCT respectively on the microscopy image dataset \cite{fmd_dataset}. Row 3 and 4 use $\sigma=30$ (out of distribution noise). Row 5 and 6 use the Poisson noise, having same standard deviation as Gaussian noise $\sigma=25$. The red line highlights the fusion value that achieves the best performance with respect to the corresponding measure (PSNR, SSIM and MSE). The $\Delta$ value is the difference to the DNN denoised output ($w=1$).}
\label{fig:ood}
\end{figure}

In this section, we evaluate the phenomena of OOD. We consider three types of OOD: images having different noise levels from the training set; images from another domain; images having different noise type. Figure \ref{fig:ood} shows the quantitative results on the OOD dataset. As we could observe, the fusion no longer achieves the best performance at $w=1$, in terms of PSNR, SSIM, and MSE. Instead, in some cases, the results using the Gaussian filter are already better compared to the results using DL. This problem of DL is often referred to as poor generalizability, i.e., DL methods only perform well on seen data. Images denoised using Gaussian, on the other hand, do not have this issue. It is capable of any kind of denoising, through filtering the high-frequency components. For the microscopy data (first two rows in Figure \ref{fig:ood_fusion}), we could find the border of the cell (first row) of the DnCNN denoised output is wrong, and some weird line patterns on the background of the second row that do not exist in the original noisy image. Also, when we increased the noise levels (middle two rows in Figure \ref{fig:ood_fusion}), the DL methods are no longer capable of removing the noise entirely. The training set is only on $\sigma=25$ and the model is incapable of predicting a residual map that has a higher absolute value, say $\sigma=35$. When facing a different noise type (last two rows in Figure \ref{fig:ood_fusion}), there are some black pots using DnCNN (fifth row) and the background of the last row is noisy.

In a nutshell, when facing OOD data the change of data distribution nullifies the learned feature maps from the training process and DL models no longer produce desirable denoised outputs. In such a case, the proposed fusion algorithm takes advantage of both the reliable filter to maintain the basic structures and the DL methods to retrieve more sharp but possibly wrong details. Allowing users to change the fusion weight offers one more layer of protection against incorrect hallucinations.

\subsection{Super resolution}

\begin{figure}[ht]
\captionsetup[subfigure]{labelformat=empty}
\begin{center}
\input{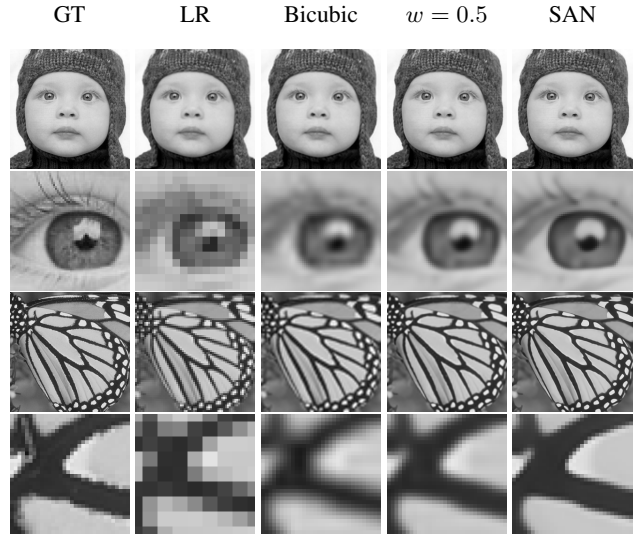}
\end{center}
\caption{Adapting the fusion methods to super-resolution tasks, without the confidence map. GT, LR stand for ground truth and low resolution respectively. The DL model used is SAN on 4x problem instances.}
\label{fig:sr_samples}
\end{figure}

In this section, we evaluate the possibility of extending our architecture to super-resolution. In fact, while reliable denoising relies on smoothing filter, reliable increase of the resolution uses interpolation methods. Most notably, bilinear are bicubic interpolation are widely used to scale an image upwards. DL methods are commonly adopted but on the other hand they are also more susceptible to the hallucinating effect, and thus run the risk of mis-interpretation of their output. We focused on low-scale 4x SR and observed that our methods could be adapted for these tasks. In particular, the reliable component could be defined by an interpolation filter. The different fusion techniques presented previously are still applicable and do not require changes, as shown in Figure \ref{fig:sr_samples}. The confidence model would have to be retrained accordingly with the new inputs, although it is possible to exploit the model trained on denoising with limited accuracy.

\section{Conclusion}\label{conclusion}
In this report, we have presented a novel framework CCID to address the interpretability and generalizability problem to the existing DL-based denoising methods. In our approach, we allow users to interpret what has been done and where the DL method might go wrong through explicit confidence map prediction. The confidence map gives users concrete feedback, helping users to understand the output of the DL model. Also, by introducing the concept of a reliable filter, we are free to vary the fusion weight between the two intermediate denoised results to select the preferred candidate. Through extensive experiments, we show that the fusion process offers both flexibility and robustness against heterogeneous inputs. Meanwhile, the same infrastructure is applicable to SISR with minor changes.

There are still a few limitations of the current approach. First, since the fusion happens in the frequency domain, it would be better to have confidence prediction in the frequency domain directly, to facilitate the confidence-guided fusion. However, due to the challenges it introduces, this idea is not implemented yet. Also, currently, the confidence prediction network is trained only on different noise levels. To have a robust prediction of the confidence in a general setting, more extensive training is required.

\clearpage
{\small
\bibliographystyle{ieee_fullname}
\bibliography{main}
}

\end{document}